# Van der Waals Epitaxy of Pulsed Laser Deposited Antimony Thin Films on Lattice-matched and Amorphous Substrates

Daniel T. Yimam,* Majid Ahmadi, Bart J. Kooi*

Zernike Institute for Advanced Materials, University of Groningen, Nijenborgh 4, 9747 AG Groningen, The Netherlands

d.t.yimam@rug.nl, b.j.kooi@rug.nl

## Abstract

Monatomic antimony thin films have recently attracted attention for applications in phase change memory, nanophotonics, and 2D materials. Although some promising results have been reported, the true potential of Sb thin films is still hindered by the scalability issue and the lack of reliable bottom-up production. Here we demonstrate the growth of Sb thin films on a lattice-matching and amorphous substrates using pulsed laser deposition (PLD). C-axis out-of-plane textured Sb thin films were successfully deposited on $Sb_2Te_3$ and $SiO_2/Si_3N_4$ substrates. In the case of growth on $Sb_2Te_3$, we show that an intermediate phase is formed at the $Sb_2Te_3$-Sb interface playing a crucial role in forming a solid coupling and thus maintaining epitaxy leading to the production of high-quality Sb thin films. A 3 - 4 nm amorphous Sb seed layer was used to induce texture and suitable surface termination for the growth of Sb thin films on amorphous substrates. The deposition parameters were fine-tuned, and the growth was monitored *in situ* by a Reflective High Energy Electron Diffraction (RHEED). Scanning/Transmission Electron Microscopy (S/TEM) unveiled the local structure of produced films showing the formation of β-phase Sb thin films. Our results demonstrate the feasibility to produce very smooth high-quality antimony thin films with uniform coverage, from few layers to large thicknesses, using pulsed laser deposition. We believe the results of our work on scalable and controllable Sb growth have the potential to open up research on phase-change materials and optoelectronics research.

## Keywords

## Introduction

The success of graphene opened a gate of new research to find two-dimensional (2D) materials with attractive properties.[1] There is great interest, due to excellent properties such as topological insulator and tunable band gap, in elemental 2D materials, including layered antimony (Sb).[2,3] The added structural stability in an ambient atmosphere, as opposed to black phosphorus (bP),[4] and the theoretically predicted large bandgap[5] gave Sb an edge against other elemental layered materials for future optoelectronic applications. The surface termination of substrates and passivated layers play a crucial role during the growth of highly textured crystals.[6–8] Surface passivation was previously used to reconstruct the Si dangling bonds by elements like Sb and Bi so that improved epitaxy can be achieved.[7,9,10] Although the passivation resolved the strain buildup from the lattice mismatch of most chalcogenides with Si,[11] the process required additional steps to remove the native oxide and reconstruct the Si surface by annealing at a high temperature. Recently it has become a common practice to directly grow epitaxial films on an amorphous substrate with a two-step method using a seed layer.[12,13] The seed layer produces a suitable surface termination for subsequent growth and particularly quintuple based materials like $Sb_2Te_3$, $Bi_2Te_3$ and $Bi_2Se_3$, also well-known as topological insulators, turned out effective as seed layers to grow films by homoepitaxy[13] and heteroepitaxy.[14–16] These quintuple based materials are also suitable substrates for depositing a few layers of antimony thin films. Previous works on the 2D – 2D growth of Sb on lattice matching Te – and Se – terminated topological insulators have shown the possibility of producing high-quality films with attractive topological surface states. [17–22]

Another avenue where elemental antimony thin films have attracted interest is for their phase-switching properties. It has been shown that the rapid crystallization of bulk Sb is avoided by confining amorphous Sb in films of only a few nm thick.[23,24] The confinement of Sb thin films increased the amorphous phase stability by retarding the structural relaxation over time.[25] Soon after stabilizing the amorphous phase, many findings emerged with applications utilizing the optical and electrical contrasts between the amorphous and crystalline phases of monatomic Sb thin films. From nanophononics and optoelectronics,[26–28] to phase change memory[23,29] and neuromorphic architectures,[24] monatomic Sb shows great promise for future applications. The functionality is extended further with an added degree of freedom, because the properties of Sb thin films depend strongly on thickness in this few nm thick range.[26,28] Although complex phase change material alloys, for example $Ge_2Sb_2Te_5$, were considered to have best properties, the "elemental" nature of monatomic Sb comes with an upper hand from the thin film production side, and phase change memory device endurance perspective. Ternary and quaternary phase change materials suffer from compositional variations and phase separation during successive phase switching. Indeed, monatomic antimony thin films will not have these problems.

Proper growth control of high-quality antimony films is of great importance, given the current interest in the material's properties for potential applications in different fields of study. Hence, here we report the van der Waals (vdW) epitaxy growth of monatomic Sb thin films on a lattice-matched substrate $Sb_2Te_3$ and on an amorphous surface ($SiO_2$) using pulsed laser deposition (PLD). So far, the majority of reported results on the growth of layered Sb employed molecular beam epitaxy (MBE),[21,30–34] and a small number of results used other techniques like exfoliation[4,35] and physical and chemical vapor deposition (PVD and CVD).[36–39] MBE is very powerful and versatile in the growth of high-quality crystals. However, the lattice-matching requirement and the need for surface reconstruction hinder the technique's applicability to a broader selection of materials and substrates.[40] Moreover, smooth Sb films showing uniform coverage over large area have not been produced using MBE.[21,30–34]

PLD provides a practical alternative with exact stoichiometry transfer, fast and flexible growth conditions, and a relatively low substrate temperature needed to produce high-quality layered crystals.[41] So far, to our knowledge, no reported result exists for the vdW epitaxy growth of monatomic Sb thin films by PLD. In addition to the vdW epitaxy of antimony on a lattice-matching $Sb_2Te_3$ seed layer, we show the possibility of producing high-quality layered Sb crystals on an amorphous substrate in a two-step growth method. The quality of Sb thin films was monitored *in situ* by a reflection high energy electron diffraction (RHEED) setup. The production of high-quality films is confirmed by atomic resolution imaging using scanning transmission electron microscopy (STEM). In addition, the full coverage and smoothness of the produced films were characterized by scanning electron microscopy (SEM) and atomic force microscopy (AFM). Our work shows substantially improved quality antimony films and thereby opens up several avenues from the production and application perspective in the fields of phase change memory, nanophononics, and optoelectronics.

## Results and discussion

*In situ* RHEED measurements can provide relevant information about thin film morphology, quality, and crystallinity during depositions.[14–16] Fig. 1 shows RHEED images captured during Sb thin film depositions on $Sb_2Te_3$ and Sb seed layers. The figures provide information about individual interfaces in the heterostructures, starting from the initial SiO$_2$ substrate to the final Sb layers. Fig. 1a presents a cloudy RHEED pattern of the starting thermal oxide substrate. The deposition starts with a 3 – 4 nm thick Sb$_2$Te$_3$ seed layer at room temperature. This first layer is amorphous, as confirmed by RHEED in Fig. 1b, since the deposition is at room temperature. The idea here is to control the out-of-plane epitaxy by the self-organizing nature of the seed layer upon heat treatment.[12] Once the seed layer is annealed to the deposition temperature, 210 °C in our case, the 3 – 4 nm thick seed layer crystallizes in a fashion with self-alignment of specific crystallographic planes parallel to the surface. Here, the initial amorphous Sb$_2$Te$_3$ layer crystallizes with the c-axis out-of-plane because the quintuples blocks with van-der-Waals (vdW) like bonding between them align parallel to the surface. Fig. 1c. shows the RHEED pattern of the seed layer after annealing to 210 °C. A sharp, streaky pattern indicates a smooth surface and high crystallinity. Note that although the seed layer induced a purely c-axis out-of-plane epitaxy, the in-plane orientation of the crystalline grains is random. Soon after the Sb$_2$Te$_3$ seed layer's crystallinity and epitaxy are confirmed, Sb growth is started at 210 °C. Fig. 1d shows the RHEED pattern after 20 nm Sb layer deposition. The streaky patterns progressed throughout the Sb thin film thickness, indicating that in addition to the surface smoothness, the c-axis out-of-plane epitaxy is maintained.

Sb$_2$Te$_3$ is carefully selected as a seed layer for Sb growth for several crucial reasons. The first reason is the similarity of the crystal structures of both Sb$_2$Te$_3$ and Sb layers. The Sb$_2$Te$_3$ quintuple layers crystallize in a trigonal crystal structure (space group R$\bar{3}$m) where the two cation Sb atoms and the three anion Te layers are stacked along the c-axis. Monatomic antimony has the same trigonal crystal structure, where bilayers of Sb are stacked along the c-axis out-of-plane orientation (see the cross-section images below in Fig. 3). Moreover, the Te – termination of the last quintuple layer in Sb$_2$Te$_3$ is ideal for the vdW epitaxy of Sb bilayers for a layer-by-layer growth mode. Another reason is the nearly perfect lattice match of the Sb$_2$Te$_3$ quintuple layers with the Sb bilayers. The structure of Sb$_2$Te$_3$ (a = 4.264 Å, and c = 30.458 Å) and Sb (a = 4.30 Å, and c = 11.22 Å) have an in-plane lattice mismatch of <1%, which can be considered as a perfect lattice match since the slight mismatch can easily be accommodated (gradually) at the vdW-like gaps in the Sb top layer similar to the strain relaxation behaviors observed between all pairwise combinations of Sb$_2$Te$_3$, Bi$_2$Te$_3$, and GeTe.[14,15]

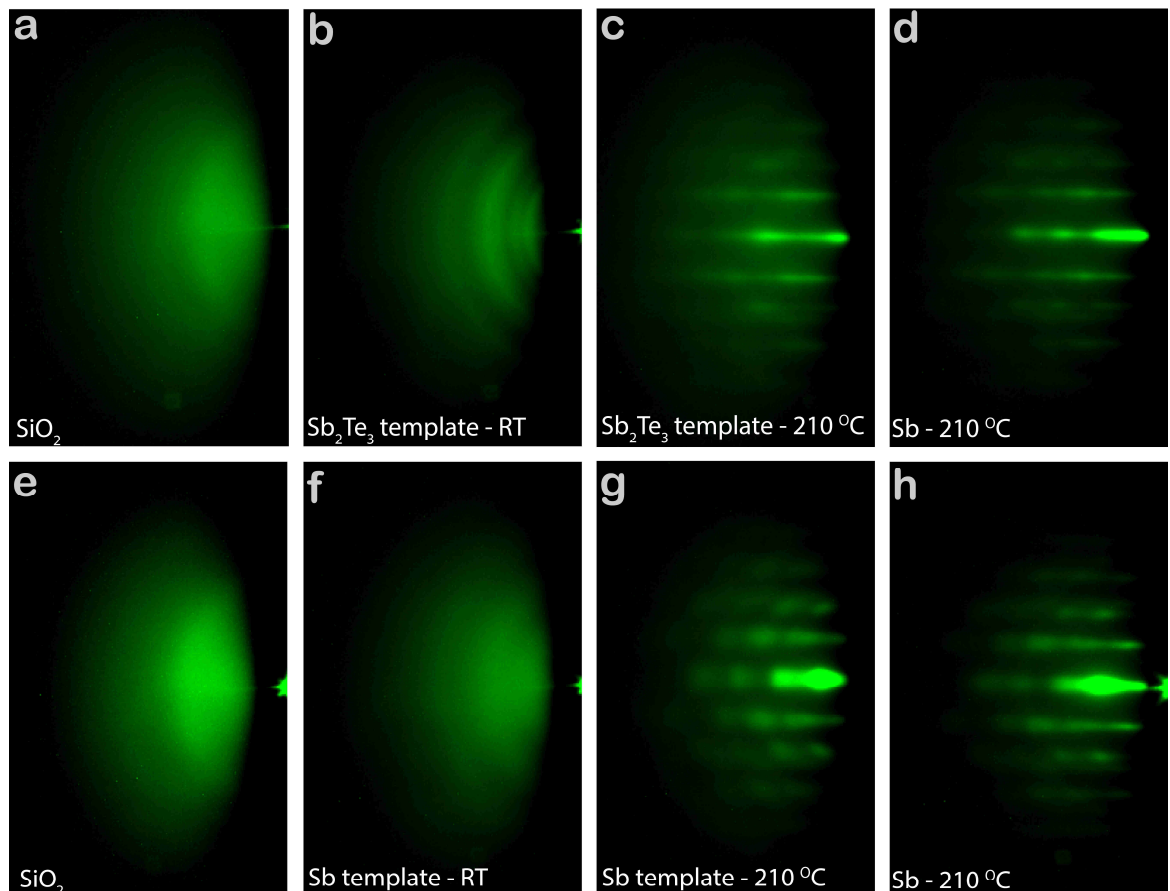

**Figure 1**. RHEED pattern images of Sb deposition on Te- and Sb- terminated seed layers of Sb$_2$Te$_3$ and Sb. In (a) and (e), the RHEED patterns of the starting thermally grown SiO$_2$ substrates are shown. The SiO$_2$ layers have a thickness of 300 nm. A few layers (3 – 4 nm thick) of (b) Sb$_2$Te$_3$ and (f) Sb seed layers were deposited on the SiO$_2$ substrate. Seed layers of Sb$_2$Te$_3$ (c) and Sb (g) were annealed to 210 °C to induce the c-axis out-of-plane texture. After annealing the seed layers, Sb growth continued. The RHEED patterns in (d) and (h) show the final film texture after Sb's heteroepitaxial and homoepitaxial growth consecutively.

The heteroepitaxy growth of Sb on the Sb$_2$Te$_3$ seed layer produced smooth and high-quality films. Therefore, the same process can be applied using directly Sb as seed layer for subsequent Sb growth. After optimization of the PLD conditions, we also in this case grow crystalline Sb thin films in which all domains have their c-axis out-of-plane, *i.e.*, with the Sb bilayers parallel to the SiO$_2$ surface. The RHEED pattern images collected during the homoepitaxy growth of Sb thin films on the Sb seed layer are presented in Fig. 1e – 1f. Here, similar to when the Sb$_2$Te$_3$ seed layer was used, the complete interface information is presented in sequence starting from the SiO2 substrate in Fig. 1e. In Fig. 1f and 1g, the Sb seed layer deposition at room temperature and after annealing to the deposition temperature of 210 °C are shown respectively. As expected, the room temperature deposited 3 – 4 nm Sb layer is amorphous (confirmed by the cloudy RHEED image), and the annealed layer is crystalline with a streaky RHEED pattern showing the c-axis out-of-plane epitaxy formation. Furthermore, as confirmed by RHEED in Fig. 1h, the c-axis out-of-plane epitaxy is maintained when growth continues at 210 °C. The homoepitaxy growth of Sb on the Sb seed layers is preferred compared to the heteroepitaxy growth of Sb on the $Sb_2Te_3$ seed layer. In addition to the 'chemical consistency,' the lack of strain from the exact lattice template is a large advantage.

Fig. 2 shows the morphology and plan view structural analysis of Sb films on the $Sb_2Te_3$ seed layer, where the heterostructure schematics are illustrated in Fig. 2a. During deposition, thermal oxide and $Si_3N_4$ membrane were used to couple the large area SEM and AFM characterizations to a local STEM characterization. Figures 2b and 2c show the images from the surface morphology characterizations by SEM and AFM, respectively. The observed morphologies provide crucial information about film quality, grain size, and surface roughness. Crystal domains of ≈ 200 nm in size are stacked together laterally. From the AFM scan in Fig. 2c, the root-mean-square (RMS) roughness of a 20 nm Sb film on a 12 nm $Sb_2Te_3$ layer is determined to be 0.9 nm indicating a smooth surface. Although PLD-grown amorphous films generally have a very smooth surface, this is not the case for crystalline films. When transferring ablated plasma plume to a substrate, multiple parameters determine the type of growth for the formed crystal. From laser energy and repetition rate to substrate temperature and processing gas pressure, parameter optimization is necessary to access the desired growth type. To achieve smooth crystalline Sb films with large grain sizes, combining a high laser repetition rate with a low Ar background pressure during deposition is required. The combined PLD optimization thus produced for a heterostructure of 12 nm $Sb_2Te_3$ and 20 nm Sb thin film a uniform surface coverage over cm sized area with low RMS roughness of ≈ 0.9 nm and with grains that have their c-axis out-of-plane and have sizes of ≈ 200 nm. Supplementary information (SI) Fig.S1 provides a detailed explanation of parameter optimization towards this layer-by-layer growth.

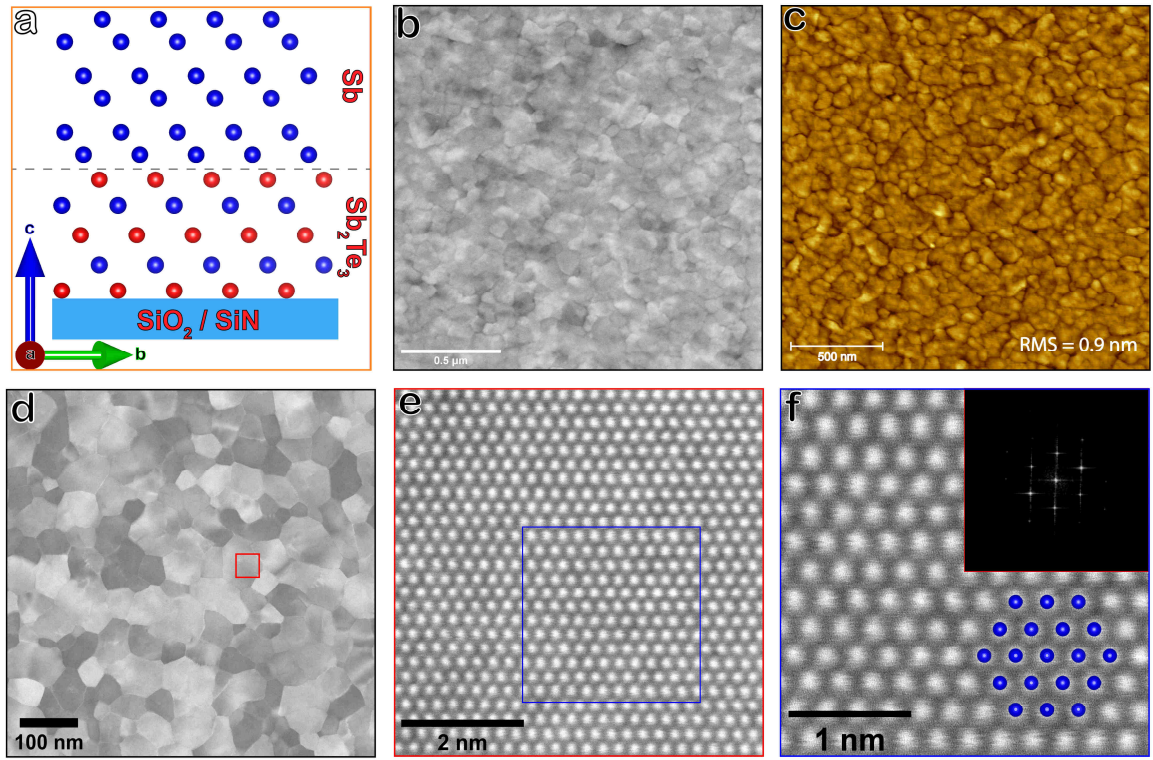

**Figure 2**. Large and small area plan-view characterization of Sb grown on $Sb_2Te_3$ seed layer. (a) A schematic diagram shows Sb's c-axis out-of-plane growth on the $Sb_2Te_3$ seed layer on amorphous substrates. (b) An SEM and (c) AFM scan images of 20 nm Sb thin film grown at 210 °C on a 12 nm $Sb_2Te_3$ layer. Crystal grains with lateral sizes of ≈ 200 nm have a sharp c-axis out-of-plane texture but with a random in-plane orientation. The root-mean-square (RMS) roughness of 0.9 nm indicates a smooth surface. (d) A large area HAADF-STEM image of Sb grown on 3–4 nm $Sb_2Te_3$ seed layer. Crystal grains of ≈ 90 nm stacked laterally, covering the entire $Si_3N_4$ substrate surface. (e) High-resolution HAADF-STEM image of one of the crystal grains seen in (d), and (f) an enlarged image of the Sb atoms with a schematic showing overlaying the hexagonal Sb lattice and the FFT in the inset.

For high-resolution HAADF-STEM imaging of Sb thin films grown on a $Sb_2Te_3$ seed layer, deposition has been done on a $Si_3N_4$ membrane. Here the $Sb_2Te_3$ layer thickness has been set to 3 – 4 nm to minimize the seed layer effect for the plan-view imaging. Fig. 2d presents a large area HAADF-STEM image showing multiple grains covering the $Si_3N_4$ substrate surface. The average grain size here is ≈ 80 – 90 nm. The reduced crystal grain size is associated with the thinner $Sb_2Te_3$ seed layer, *i.e.* the lateral domain size increases when the $Sb_2Te_3$ film growth is continued at 210 °C. As a result, larger crystals have been deposited with a thicker $Sb_2Te_3$ bottom layer of 12 nm in thickness observable in the cross-section image of a sample with about 20 nm Sb film on top (see Fig. 3a). A high-resolution atomic image of a specific Sb crystal is provided in Fig. 2e, with an enlarged image given in Fig. 2f. A clear hexagonal periodicity corresponds to the [0001] zone axis of β-phase antimony.

A cross-section atomic-resolution HAADF-STEM image of the $Sb_2Te_3$–Sb heterostructure is shown in Fig. 3a, with the enlarged part in Fig. 3b. The cross-section image in Fig. 3a shows sharp interfaces between the $SiO_2$ substrate, the stacked quintuple $Sb_2Te_3$ layers, and the Sb bilayers. In HAADF-STEM mode, the atomic columns are always bright spots in a dark surrounding, and the brightness of the spots scales with the average atomic number Z of the columns. Due to the closeness of Sb and Te atoms in the periodic table, their Z contrast is (unfortunately) negligible. Still, the $Sb_2Te_3$ quintuples (Te-Sb-Te-Sb-Te units) and the Sb bilayers can be distinguished readily in Fig. 2a and 2b. Indeed, the layers are aligned parallel to the substrate surface, and the Sb grows epitaxially on the $Sb_2Te_3$. STEM-EDX elemental mapping of the Sb – $Sb_2Te_3$ heterostructure is presented in Fig. 2c-2f. From the elemental mapping results, the Sb and $Sb_2Te_3$ layers are identifiable, and it can be assured that there is a uniform Sb layer on top of the $Sb_2Te_3$ layer.

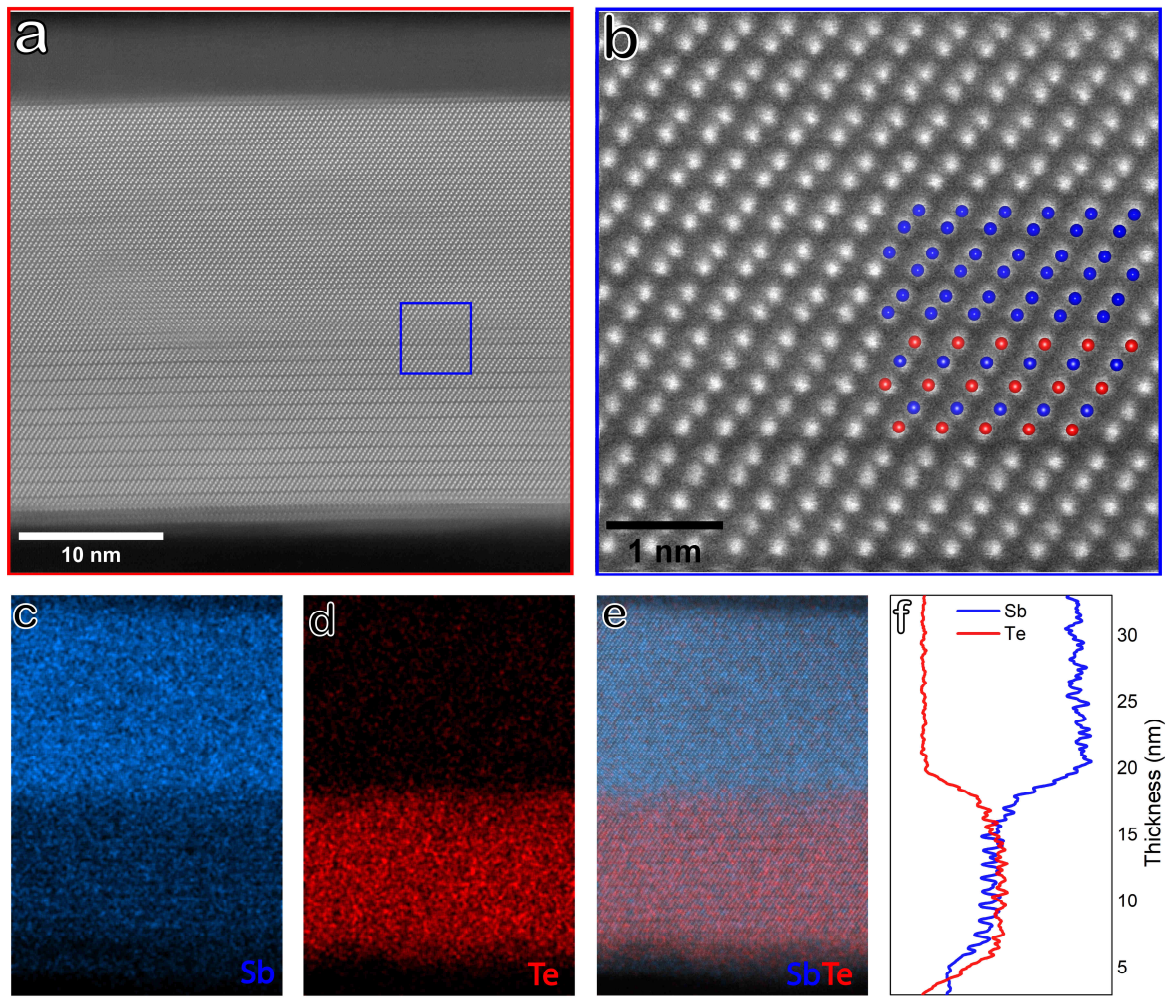

**Figure 3**. Cross-section STEM results for an Sb film (≈ 20 nm thick) on a ≈ 12 nm thick $Sb_2Te_3$ layer. The deposition started with a 3 – 4 nm $Sb_2Te_3$ seed layer on a silicon wafer covered with thermal oxide. The growth direction in all images is from bottom to top. (a) STEM – HAADF

images show a complete overview of the whole layer containing the $Sb_2Te_3$ bottom layer and the Sb top layer. (b) Enlarged part of (a) showing the atomic structure and epitaxy of Sb on $Sb_2Te_3$ more clearly. In (c) and (d) STEM-EDX color map (based on spectrum imaging, *i.e.*, for each pixel an EDX spectrum is recorded) showing the distribution of Sb in blue and Te in red. (e) A combined (Sb and Te) elemental mapping with clear color contrast near the $Sb_2Te_3$–Sb interface, and in (f) STEM_EDX line profiles for Sb (blue line) and Te (red line).

The HAADF-STEM images of the $Sb_2Te_3$–Sb cross-section show the lattice-matched epitaxy, where the $Sb_2Te_3$ layer is a suitable template for Sb growth. However, a more intricate interesting effect can be observed at the interface, where the last $Sb_2Te_3$ quintuple layer and the first Sb bilayer couple (to form a new phase), see Fig. 4. The last distinctly observable $Sb_2Te_3$ quintuple occurs at the lower blue box, and the first clearly observable Sb bilayer occurs at the upper blue box in Fig. 4a and in the magnified image of the interface in Fig. 4b. In-between, inside the overlayed red box, there is a vdW block of seven atomic layers which can be typified as a $Sb_4Te_3$ layer (*i.e.*, Te-Sb-Te-Sb-Te-Sb-Sb units).[42,43] So, the actual position of the interfaces is indicated by the interfaces of the red and blue overlayed boxes. However, the interface is not formed by a similar width vdW-like bonding, such as between the $Sb_2Te_3$ quintuples or between the Sb bilayers, but by a shorter and thus stronger (more covalent/metavalent) bond. These distances can be readily quantified based on direct measurements in the line scans presented at the left side of Fig. 4a and the right side of Fig. 4b. The vdW-like gaps between the $Sb_2Te_3$ have a size of 278±4 pm. Similarly, the vdW-like gap between the Sb bilayer is 244±4 pm. However, at the interface between $Sb_2Te_3$ and Sb, *i.e.* within the $Sb_4Te_3$ layer, the gap between the last Te-plane and the first Sb plane on top is 219±4 pm. So, this is even 10% smaller than the vdW-like gap between the bilayers in Sb.

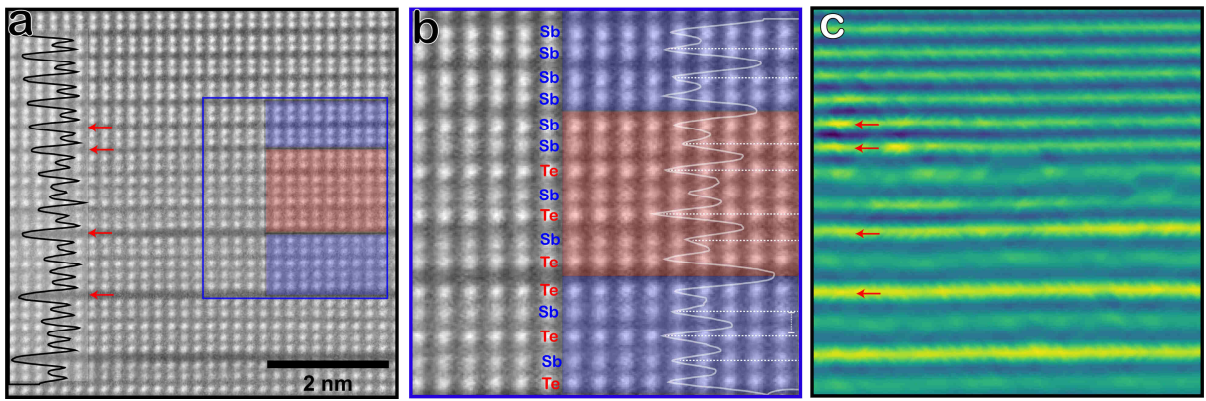

**Figure 4**. A new phase of $Sb_4Te_3$ is formed at the $Sb_2Te_3$–Sb interface. (a) Atomic resolution HAADF-STEM image of the Sb-$Sb_2Te_3$ interface with the line profile overlayed on the left. Lower blue box on the right shows the $Sb_2Te_3$ layer on the bottom, upper blue box the Sb bilayers on the top, and the red box the intermediate new layer of seven atomic planes of Sb$_4$Te$_3$. (b) An enlarged image of (a), with the line profile on the right. (c) The color map of the monolayer distance difference from the Atomap software tool. The large distance differences, *i.e.*, the vdW-like gaps, are bright yellow. At the interface between the last Sb$_2$Te$_3$ and the first Sb, within the Sb$_4$Te$_3$, the vdW-like gap disappears.

Another piece of evidence for the presence of a new phase at the Sb$_2$Te$_3$–Sb interface comes from the high-resolution image analysis of Fig. 4a using the software tool Atomap.[44] Here, the interplanar distance differences in vertical direction (parallel to c-axes of Sb$_2$Te$_3$ and Sb) are extracted and plotted in Fig. 4c after the position of each atomic column is accurately determined and horizontal atomic planes are constructed. The color contrast gives the interplanar distance differences, yellow being the largest and blue the smallest. As indicated by the red arrows, the vdW-like gaps are readily visible with the highest distance difference as they should be. The distance difference also holds for the gaps in the bilayers of Sb atoms, alternating long and short distances between atomic layers. When looking closer to the interface, the last Sb$_2$Te$_3$ and the first Sb, we notice the color contrast vanishes, indicating a reduced distance difference. Therefore, strong evidence is provided that a new phase, Sb$_4$Te$_3$, is formed at the interface with relatively strong coupling between the last Sb$_2$Te$_3$ and the first Sb bilayer on top, such that an excellent epitaxy of Sb on Sb$_2$Te$_3$ is achieved.

The more important novelty of this work, in addition to the layer-by-layer growth of Sb on a lattice-matched Sb$_2$Te$_3$ template layer, is the homoepitaxy growth of Sb. Instead of Sb$_2$Te$_3$, also Sb can be used as a seed layer, and then pure and highly textured Sb films can be grown. Again, films uniformly covering the substrate can be produced where all domains have their c-axis out-of-plane. Similar to the Sb$_2$Te$_3$ seed layer, the deposition starts with a 3 – 4 nm amorphous Sb thin film deposited on the SiO$_2$/Si and Si$_3$N$_4$ substrates at room temperature. In our previous work, we showed how accurately we could produce smooth amorphous Sb thin films with thicknesses ranging between 2.7 – 6.0 nm with complete surface coverage over large areas.[28] Once the sample is annealed to the deposition temperature of 210 ºC, the amorphous Sb layer crystallizes in the trigonal crystal structure (the same space group as Sb$_2$Te$_3$) with c-axis out-of-plane orientation where Sb bilayers are stacked together with a vdW-like gap between them. After the c-axis epitaxy is induced in the seed layer, Sb deposition is continued at 210 ºC. The layer-by-layer growth will then follow the initial Sb seed layer lattice. The RHEED pattern images in Fig. 1e – 1h show the deposition process of Sb on the Sb seed layer. After deposition, the morphology and structure of the Sb thin films were investigated by HAADF-STEM. A plan-view image of Sb thin film prepared on a Si$_3$N$_4$ substrate is shown in Fig. 5, with the homoepitaxy ' heterostructure' schematically depicted in Fig. 5b.

A homogeneous coverage of Sb thin films with c-axis out-of-plane orientation is seen in Fig. 5a and the inset FFT pattern. Here, similar to Sb thin films deposited on the $Sb_2Te_3$ seed layer, the in-plane randomly oriented crystal grains are stacked together laterally without openings, creating a smooth surface (see supplementary information (SI) Fig. S3). However, the lateral sizes of the domains are substantially smaller in this case, only 15 - 20 nm, and thus remain similar to the film thickness. It is worth mentioning here that the linear correlation between laser pulse number with thickness means we can also produce Sb films ranging from a few bilayers to larger thicknesses.

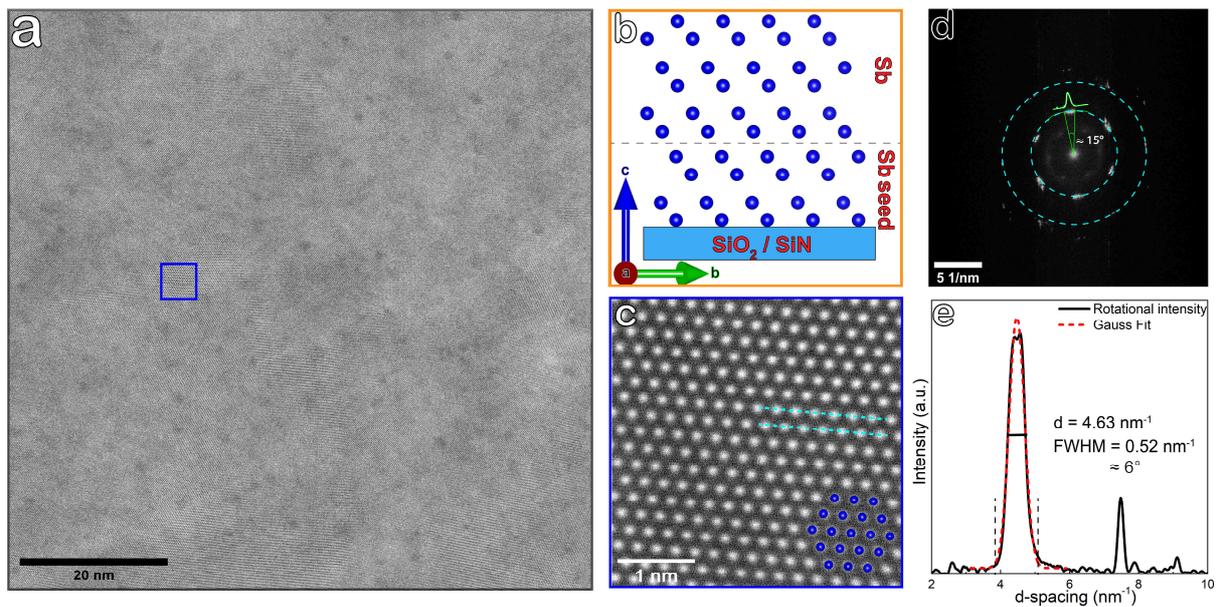

**Figure 5**. High-resolution plan-view imaging of Sb thin films produced using an Sb seed layer. A large area view of the film surface, with multiple Sb crystal grains, is shown in (a). No opening between grains and the FFT in (d) confirms the c-axis out-of-plane texture. The FFT spots in (d) have an angular spread of ≈ 15 degrees between all crystal grains. The schematic for the homoepitaxy growth of Sb is presented in (b). Here, the first 3 – 4 nm thick Sb layer was first used as a seed layer to promote texture. Atomic resolution imaging of a single Sb crystal is shown in (c). The overlaying hexagonal Sb lattice confirms the production of β-phase Sb thin films. In (e), the average rotational intensity of the FFT pattern shown in (d) is presented. The central peak corresponds to a d-spacing of 4.63 $nm^{-1}$, and from the central peak's Gauss fit, the FFT spots spread is expressed with a FWHM value of 0.52 $nm^{-1}$ corresponding to ≈ 6 degrees spread.

From the FFT pattern in Fig. 5d, all grains within the field of view of Fig. 5a are highly c-axis out-of-plane oriented, and the in-plane orientation is limited to a slight angle rotation (≈ 15 degrees for all). The limited a- b- axis rotation is evident in the FFT pattern where we would have expected circular rings as seen in the supplementary information (SI) Fig.S4 for random in-plane orientation, but here the spots do not makeup rings. In turn, diffraction spots from individual grains are close to each other with a slight orientation change. Therefore, we can extract the approximate spread angle for most crystal grains from the average rotational intensity and the FWHM of the intensity peak. As illustrated in Fig. 5e, the start and end of the central intensity peak, ≈ 3.83 nm$^{-1}$ and ≈ 5.14 nm$^{-1}$, with a difference of ≈ 1.31 nm$^{-1}$, correlates to the ≈ 15 degrees measured in Fig. 5d. From the Gauss fit of the intensity peak in Fig. 5e, the FWHM value of ≈ 0.52 nm$^{-1}$ is extracted, which correlate with an average rotation angle of ≈ 6 degrees. Note that this is indeed dependent on the analyzed area where the in-plane orientation change is limited for small areas (for example, for an area in Fig. 5a), and the orientation is completely random for large areas (as seen in the supplementary information (SI) Fig.S4). Finally, high-resolution atomic imaging of a single crystal grain is presented in Fig. 5c. Similar to what we imaged for Sb grown on $Sb_2Te_3$, the hexagonal periodicity is seen, confirming the growth of trigonal β-phase antimony.

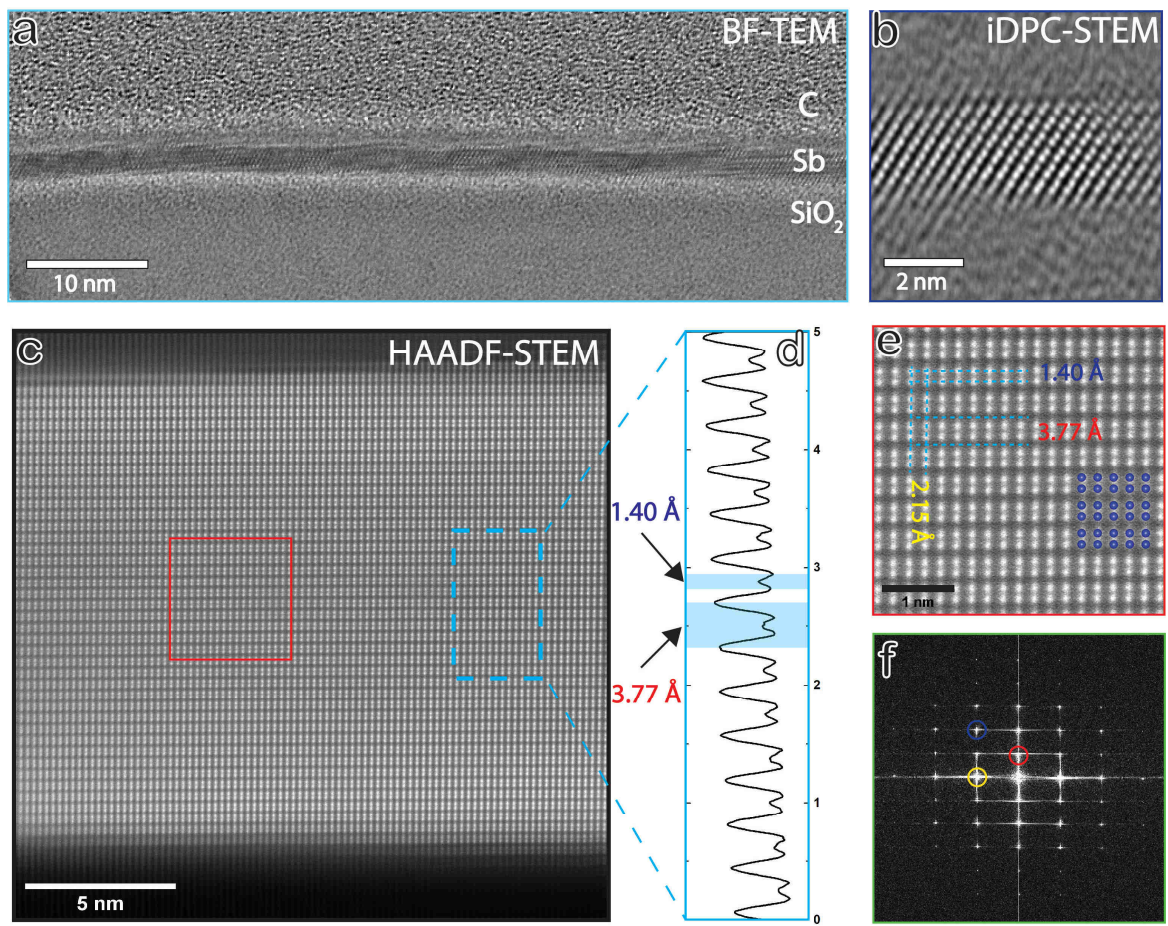

**Figure 6**. Sb thin films were grown using PLD, and a thin seed layer technique was used such that the Sb bilayers were everywhere parallel to the $SiO_2$ substrate surface. (a) BF-TEM image of a few layers of Sb/antimonene on $SiO_2$ substrate showing full surface coverage. (b) The iDPC-STEM image of a single crystal of Sb/antimonene layers with c-axis out-of-plane orientation. (c) Overview HAADF-STEM image showing the entire thickness of about 20 nm of the Sb film. The bottom $SiO_2$ substrate and the top carbon capping layer are readily visible. (d) An intensity line profile for the indicated blue dashed region allowing quantifying

interplanar distances and vdW-like gaps and (e) a magnified view of an area indicated by the red box in (c). (f) the FFT pattern was extracted from (e).

As previously mentioned, the Sb growth on amorphous substrates starts with a few layers of Sb (antimonene) with a preferred c-axis out-of-plane orientation. Fig. 6a shows a cross-sectional BF-TEM image of the few layers of antimonene grown on $SiO_2$ substrate. The layer has a thickness of only 3 – 4 nm. Starting from an amorphous phase and subsequently crystallizing the thin layer induced a crystalline phase with a preferred of c-axis out-of-plane orientation. This orientation is visible in Fig. 6b where an iDPC-STEM image shows the Sb bilayers stacked together when viewed in $[11\bar{2}0]$ zone axis. More details about the few layers of antimonene production and characterization is provided in the supplementary information (SI2). The few layers of antimonene formed on a $SiO_2$ substrate have a complete full coverage and an extremely smooth surface, suitable for use as a template. Fig. 6c shows a cross-sectional HAADF-STEM image of homoepitaxially grown Sb film on $SiO_2$ substrate. Here, the layered Sb with c-axis out-of-plane orientation (as viewed in $[10\bar{1}0]$ zone axis orientation) sandwiched between the bottom $SiO_2$ layer and the top carbon capping layer can be observed with clearly isolated interfaces. The crystalline Sb film has an overall thickness of about 20 nm. A magnified view of an area indicated by a red box is shown in Fig. 6e. The STEM images show a vdW-like gap separating the Sb bilayers. In Fig. 6d, an intensity line profile is plotted for the blue dashed rectangular area indicated in Fig. 6c. Blue overlay boxes in Fig. 6d distinguish the separation for Sb-Sb bilayers and the vdW-like gaps. We measured an interplanar separation distance of ≈1.40 Å between the Sb atoms within the bilayer and a distance of ≈3.77 Å for the vdW-like separation between the bilayers (*i.e.* the vdW gap itself is ≈2.37 Å which is close to the 2.44 Å measured above for Sb bilayers on $Sb_2Te_3$).

Our work demonstrated the possibility of producing high-quality Sb thin films using PLD. By fine-tuning the deposition parameters, the c-axis out-of-plane growth of Sb thin films was achieved on lattice matching $Sb_2Te_3$ and amorphous $SiO_2/Si_3N_4$ substrates. The quality and texture of the deposited thin films were monitored *in situ* by RHEED and ex-situ by high-resolution electron microscopy. The structural analysis of the thin films revealed the production of the most stable allotrope β-phase antimony with a trigonal crystal structure. Previously reported works, where MBE and PVD was employed (since no reported work exists for Sb thin film production using PLD), show the difficulty of producing smooth, high-quality Sb thin films. The reported results mainly show Sb islands and flakes on crystalline substrates like graphene and sapphire.[30,31,36–38,45,46] In contrast, our Sb thin films show smooth thin films with continuous and complete substrate coverage. This can be seen from the SEM, AFM, and HAADF-STEM results presented in this work. Even more interesting is our ability to produce purely c-axis out-of-plane Sb thin films on any amorphous substrate. The thin films can also be relatively small in thickness with only few layers of Sb stacked as seen in Fig. 6a and 6b. Fig. S2a also show the initial amorphous seed layer which has a thickness of about 3 nm. In addition, we previously reported our ability to produce high-quality Sb amorphous thin films with full substrate coverage and smooth surfaces. Fig. 2b and Fig. S2 of Ref [28] clearly show the high-quality Sb thin films of relatively small thicknesses.

## Conclusions

In summary, our work shows the Van der Waals epitaxial growth of antimony thin films on a lattice-matched seed layer of $Sb_2Te_3$ and on amorphous $SiO_2$ and $Si_3N_4$ substrates using PLD. We show that extremely smooth and continuous Sb thin films with full substrate coverage can be produced by precisely controlling and fine-tuning the deposition parameters. The homo- and hetero- epitaxial growth of Sb thin films, starting from an amorphous seed layer and subsequently crystallized in a preferential trigonal structure, was confirmed to have a sharp c-axis out-of-plane texture. The texture was unveiled by RHEED and local atomic resolution HAADF-STEM imaging. In addition, from the high-resolution atomic imaging analysis, we confirmed the formation of an intermediate phase on the $Sb-Sb_2Te_3$ tie line where the first Sb bilayer couples with the last $Sb_2Te_3$ quintuple during Sb deposition on $Sb_2Te_3$. The intermediate phase formation promoted an Sb-terminated surface, leading to a 2D growth of Sb with excellent epitaxy. Given the current interest in monatomic Sb thin films, our reported results of smooth and high-quality Sb thin films produced by PLD will stimulate further study on phase change materials and optoelectronics fields. Moreover, the ability to produce full coverage films with preferential orientation and texture control on any substrate will find new and exciting applications for monatomic Sb thin films and other layered materials.

## Experimental methods

**Sample preparation and thin film growth:** For $Sb_2Te_3$–Sb heterostructure depositions, a seed layer of 200 pulses (3 – 4 nm thick) of $Sb_2Te_3$ was deposited at room temperature. After the deposition, the sample was annealed to 210 °C to crystallize and to enable epitaxy with c-axis out-of-plane. The deposition of $Sb_2Te_3$ is then continued at 210 °C. After the $Sb_2Te_3$ deposition, the process is continued with Sb deposition. All depositions were done with a PLD system using KrF excimer laser (wavelength of 248 nm). The deposition parameters were optimized beforehand by varying them during extensive tests across parameter space. A laser fluence of 0.8 J cm$^{-2}$, a repetition rate of 1 Hz, and a processing gas (Argon) pressure of 0.12 mBar were used for Sb$_2$Te$_3$ depositions. Furthermore, a laser fluence of 1.5 J cm$^{-2}$, a repetition rate of 3 Hz, and a processing gas pressure of 10$^{-3}$ mBar were used for Sb depositions. After every deposition, the samples were capped by a 5 – 10 nm thick LaAlO$_x$ (with x close to 3) layer to prevent surface oxidation and chemical contamination. Before every deposition, the Si/SiO$_2$ substrates were chemically cleaned. The cleaning steps include sonication in isopropanol for 20 minutes and flushing with ethanol. The film growth and epitaxy were monitored by an *in situ* reflection high-energy electron diffraction (RHEED) setup with an accelerating voltage of 30 kV. The grazing incident (2 - 3 degrees normal to the substrate) electron beam provides information on the top few atomic layers of the deposited film. RHEED images were collected every 300 ms during depositions.

**Morphological characterization by SEM and AFM**: For morphological and structural characterizations, a set of samples were prepared on Si/SiO$_2$ substrates and Si$_3$N$_4$ TEM membranes. Scanning electron microscopy (FEI NovaNanoSEM 650 and FEI Helios G4 CX) was used to characterize the surface morphology of the deposited thin films. In addition, atomic force microscopy (AFM) images were collected by a Bruker Multimode 8 instrument, and the images were analyzed by the Gwyddion software.

**FIB sample preparation:** The FEI Helios G4 CX dual-beam SEM-FIB (scanning electron microscope)-(focused ion beam) was used to prepare an e-transparent sample. First, the protective EBID carbon(C) and platinum(Pt) were deposited on the film before IBID Pt protective layer. Next, cross-sectional chunks (dimensions: 15 x 2.0 x 5 μm$^3$) were made and transferred using easylift$^{TM}$ needle to the copper half-grid. Then the sample was thinned down to 80-100 nm thickness using standard Ga-beam processing at 30 kV with an opening window (6.0 x 5.0 μm$^2$). The remaining chunk is left thick to have a rigid frame to minimize the bending and stress released in the e-transparent window. Finally, several low kV cleaning steps (5 and 2 kV) were used to clean the side surfaces of the lamella.

**Scanning/Transmission Electron Microscopy (STEM):** The FIB prepared sample and loaded to the TEM grid was transferred to the TEM using a dedicated double tilt TEM holder optimized to collect x-rays in the TEM. The microstructure imaging of plan-view and cross-sectional layered antimony films were examined with a double-corrected and monochromated Themis Z scanning transmission electron microscope (Thermo Fisher Scientific) operating at 300 kV and equipped through high-angle annular dark-field (HAADF) STEM and integrated differential phase contrast (iDPC) STEM using ~80% coverage of a 4-segmented DF detector. The beam convergence angle was measured ~24.0, and ~35mrad mrad (for the thin sample with small grains) and the probe current of 20 pA was used for STEM imaging. Energy dispersive X-ray spectroscopy (STEM EDS maps) results were achieved with a Dual X EDS system (Bruker) using two large area detectors, capturing 1.76 steradians with a probe current of 100 pA. Data acquisition and analysis were done using Velox software.

## Acknowledgments

This project has received funding from the European Union's Horizon 2020 Research and Innovation Programme "BeforeHand" (Boosting Performance of Phase Change Devices by Hetero- and Nanostructure Material Design)" under Grant Agreement No. 824957. We thank Prof. Tamalika Banerjee, Hans de Vries, and Job van Rijn for the technical assistance of the pulsed laser deposition setup. We would also like to acknowledge support from The Zernike institute for advanced materials (ZIAM) and Groningen cognitive systems and materials center (CogniGron) to the ZIAM electron microscopy facility at the University of Groningen (RUG).

**Supporting Information Available:** Figure S1 - Surface morphologies of pulsed laser deposited (PLD) Sb thin films obtained using different deposition parameters; Figure S2 - Few layers of Sb (antimonene) production and characterization; Figure S3 - AFM scans and image analysis of $Sb_2Te_3$ – Sb heterostructure for thickness measurement, surface roughness, and depth profile across individual grains; Figure S4 - AFM and image analysis results for Sb thin film grown on the Sb seed layer; Figure S5 - Plan-view (S)TEM images of crystal grains in Sb films with a thickness of ≈20 nm on (a) $Sb_2Te_3$ seed layer (heteroepitaxy) and (b) Sb seed layer (homoepitaxy) on a $Si_3N_4$ membrane.

Supplementary Information for

**Van der Waals Epitaxy of Pulsed Laser Deposited Antimony Thin Films on Lattice-matched and Amorphous Substrates**

*Daniel T. Yimam\*, Majid Ahmadi, Bart J. Kooi\**

Zernike Institute for Advanced Materials, University of Groningen, Nijenborgh 4, 9747 AG Groningen, The Netherlands

\*Corresponding authors. Email: d.t.yimam@rug.nl, b.j.kooi@rug.nl

# SI 1 – Parameter optimizations to high-quality Sb thin films

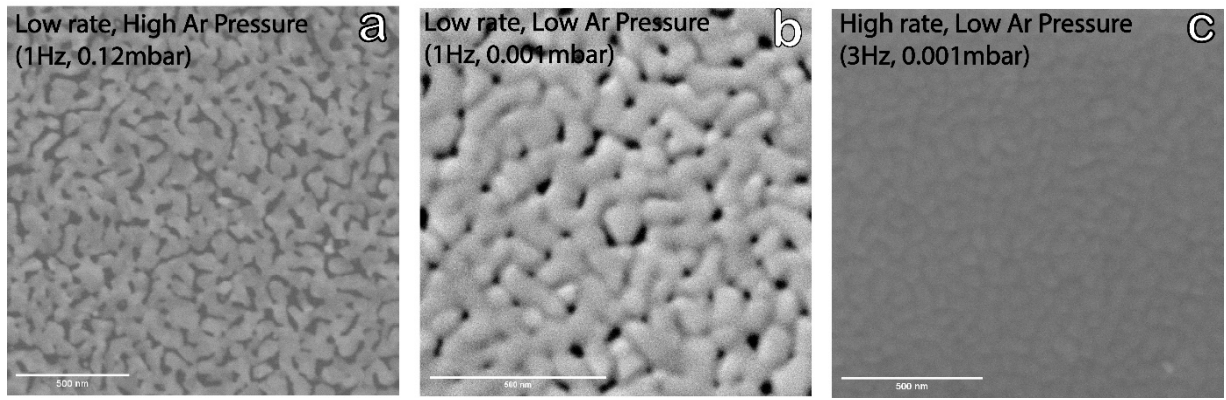

**Figure S1**. Surface morphologies of pulsed laser deposited (PLD) Sb thin films obtained using different deposition parameters. (a) Deposition at a relatively high processing Ar gas pressure and with a low laser repetition rate. This combination means low yield for the deposition and formation of islands. (b) Deposition at a low rate but also low processing Ar pressure. Here the yield is increased, but since the repetition rate is low, the island formation is still preferential over a uniform thin film formation. (c) When increasing the repetition rate while reducing the processing Ar gas pressure, smooth films with uniform coverage can be produced.

Pulsed laser deposition produces high-quality thin films with excellent control and flexibility. However, initial deposition parameter optimization must be performed to obtain these high quality thin films. Even with good starting parameters for processing gas Ar pressure and laser fluence energy from previous depositions of similar materials, optimization can take considerable time. It is even complicated when one wants to deposit a material with high surface energy and vapour pressure like pure antimony. On the one hand, the deposition parameters must be suitable to break through the material's tendency to form islands instead of a continuous film on the substrate. On the other hand, the deposition temperature has to be within a specific range to reduce evaporation. PLD offers an advantage over other deposition techniques due to the thermalized and kinetically energized large atomic species generated from the laser–target interaction. Moreover, since the amount of energized atomic species can be precisely controlled by the laser energy and the processing gas pressure, proper nucleation behaviour can be achieved on the substrate surfaces.

Fig. S1 shows some part of the parameter optimization for Sb thin film depositions. Different surface morphologies are visible for different parameter values. Here the laser fluence energy is fixed to 1.5 J cm$^{-2}$ and the substrate temperature to 210 ºC. By changing the processing Ar gas pressure and the laser repetition rate, the flux of atomic species reaching the substrate can also be changed. The laser repetition rate is fixed in Fig. S1-a and S1-b, but the processing gas pressure was changed. Island formation was inevitable for low repetition rate and high Ar pressure since the parameters are not ideal for creating high yield. However, the yield can be increased by reducing the processing Ar gas pressure, but this alone was not enough to avoid island formation. Thus, a reduction in processing Ar gas pressure had to be coupled with an increased repetition rate. This combination produced a supersaturation of atomic species on the substrate surface, thus leading to full coverage and smooth films.

**SI 2 – Few layers antimony (antimonene) production and characterization**

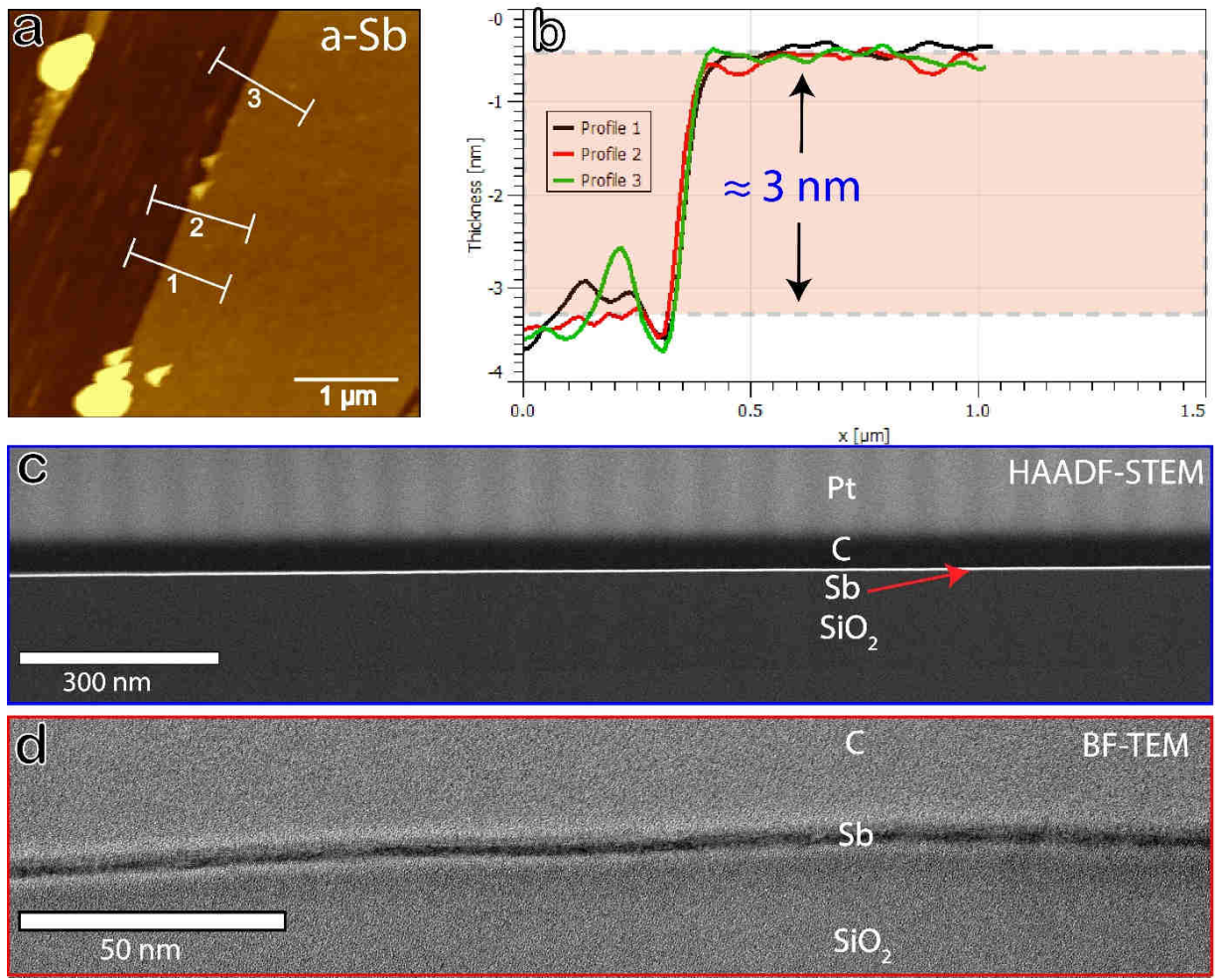

**Figure S2.** Few layers of Sb (antimonene) production and characterization. (a) An AFM scan over a scratched surface for thickness measurement. Initially, an amorphous Sb thin film was deposited. A continuous and smooth amorphous layer is visible. (b) The line profile of a step edge is created by scratching the surface. A thickness of ≈3 nm is extracted from the line profile, consistent with the expected value. (c) HAADF-STEM and (d) BF-TEM images of a cross-sectional few layers of Sb/antimonene on a SiO$_2$ substrate.

The main text explains that the Sb growth on amorphous $SiO_2/Si_3N_4$ substrates starts with a 3 – 4 nm amorphous Sb layer. An example of a deposited amorphous layer is observable in Fig. S2a where the step edge from scratching shows, as confirmed by the line profiles in Fig. S2b, a thickness value of ≈3 nm. Important to note is that the secret of creating a very smooth crystalline Sb thin film with uniform coverage on any amorphous substrate depends on the initial seed layer quality. Since extensive optimization has been done initially, a smooth amorphous Sb layer can be created and crystallized to form a crystalline template with a sharp c-axis out-of-plane texture. Figures S2c and S2d show HAADF-STEM and BF-TEM cross-sectional images of the few Sb/antimonene layers on $SiO_2$ substrate and capped by an amorphous carbon layer. The AFM and the cross-sectional images both show a few layers of Sb/antimonene with full substrate coverage.

**SI 3 – AFM characterization of crystalline Sb thin films**

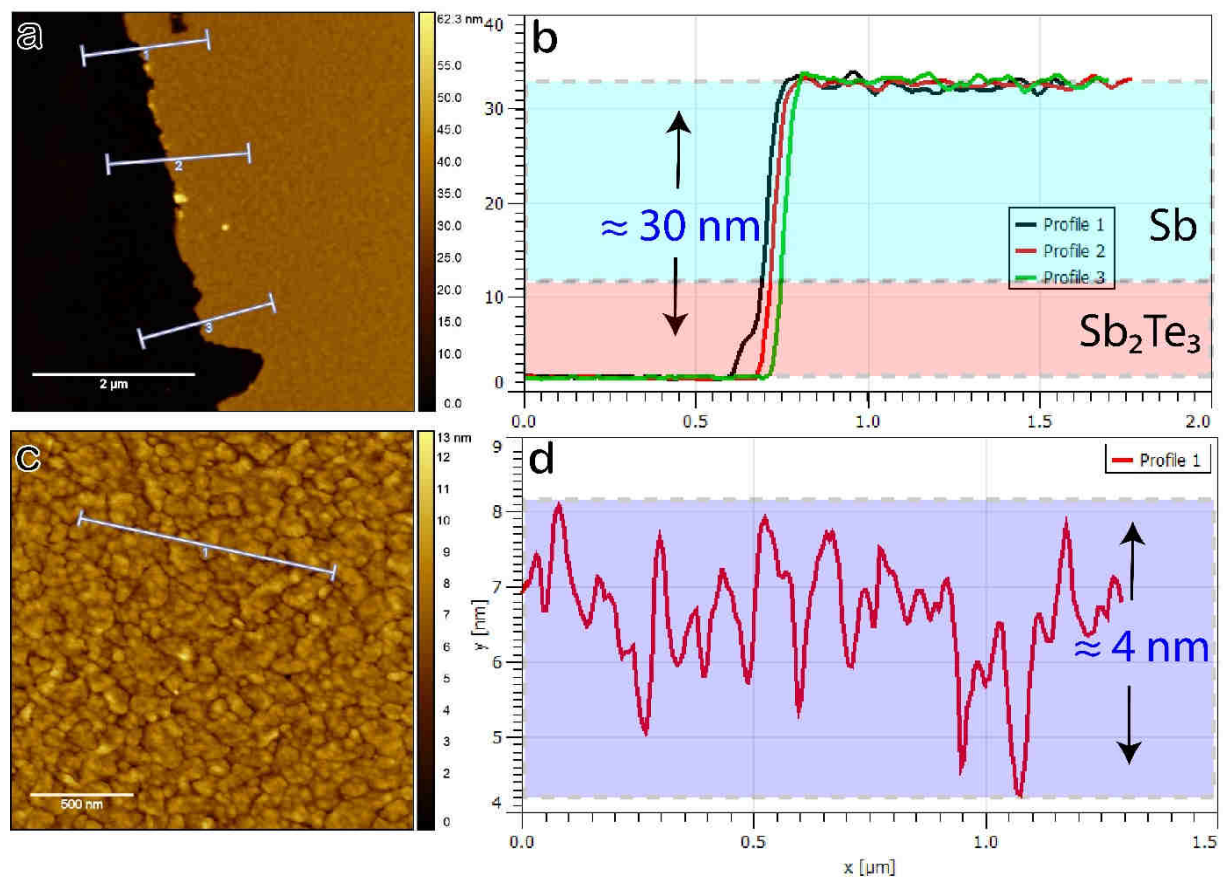

**Figure S3**. AFM scans and image analysis of $Sb_2Te_3$–Sb heterostructure for thickness measurement, surface roughness, and depth profile across individual grains. (a) Thickness measurement of the heterostructure. Here a 10 nm $Sb_2Te_3$ layer was produced, and then a 20 nm Sb layer was deposited on top. (b) The line profile shows the overall thickness of the produced heterostructure. A combined thickness of ≈30 nm is extracted, which matches well

with the intended thickness. (c) The surface morphology of the heterostructure. An RMS roughness value of ≈0.9 nm holds in this case, indicating an extremely smooth surface. (d) A profile is extracted from a line drawn randomly over the surface. Here a maximum (peak-to-peak) height difference of ≈4 nm is extracted.

The surface morphologies of Sb thin films on $Sb_2Te_3$ and Sb seed layers have been investigated using AFM. Although high-resolution HAADF-STEM imaging has produced excellent results locally, the morphological evolution of the film surface can be better studied by AFM. Here, the surface roughness and the film thickness can accurately be determined, and the separation between individual grains can be found. The grain separation and valley formation information is crucial here since the crystal grains' random in-plane orientation means complex (high angle) grain boundaries between neighbouring grains. In Fig. S3 and S4, the analysis from the AFM scans of Sb on $Sb_2Te_3$ and Sb seeds is presented. The analysis consists of thickness measurements from scratched surfaces and roughness and valley formation from line profiles of the surface morphologies.

The first sample is a heterostructure of Sb-$Sb_2Te_3$ with thicknesses of 10 nm $Sb_2Te_3$ and 20 nm Sb on top. The AFM scan over the step edge, created by scratching the surface, and the associated line profiles provide an approximate thickness of 30 nm, consistent with the expected values as shown in Fig. S3b. Fig. S3c presents the surface morphology of the heterostructure captured in tapping mode. Larger grains of Sb can easily be identified in the figure also presented in the SEM and STEM images of the main text (cf. Fig. 2). The RMS roughness of the film was extracted to be ≈0.9 nm, indicating an extremely smooth surface. The AFM imaging and analysis were also done for Sb thin films directly grown on $SiO_2$ substrates. The details of few Sb/antimonene layers production and characterization is given in SI 2. Once the crystallization of the seed layer induced a texture, growth continued and thicker samples of Sb thin films can be produced. Fig. S4a – S4d show the surface morphologies, the thickness measurement from the step edge, and the line profile. A 20 nm thick crystalline Sb thin film with c-axis out-of-plane orientation is visible in the AFM images.

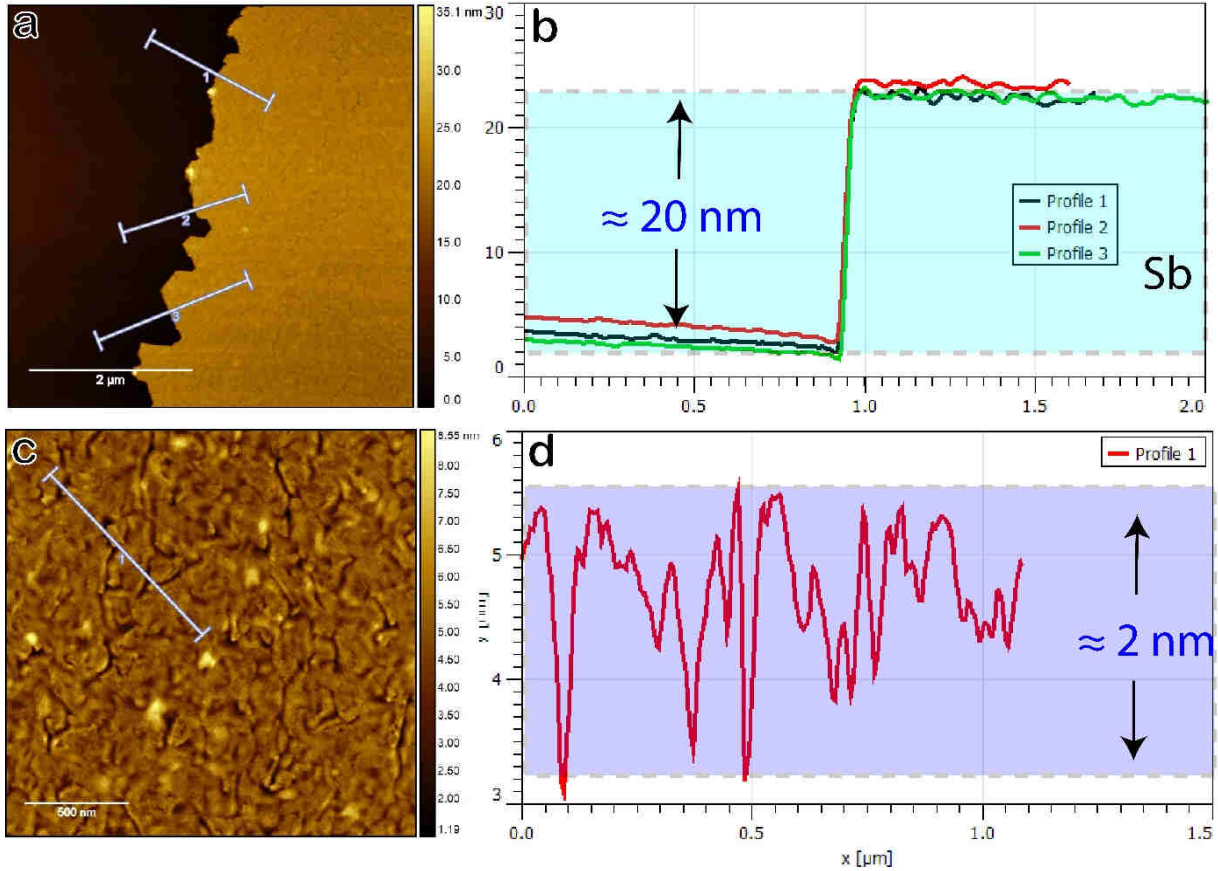

**Figure S4**. AFM and image analysis results for Sb thin film grown on the Sb seed layer. (c) After crystallizing the seed layer, growth continued, resulting in a final thickness of 20 nm crystalline Sb thin film. (d) The line profile over the scratched surface, creating a step edge, provides the thickness of the thin film. (e) The surface morphology of the final film is presented here. An extremely smooth thin film with an RMS value of ≈0.6 nm is seen. (f) A line profile, randomly drawn over the surface, gives an estimate of the maximum valley depth formed between two neighbouring grains. A maximum (peak-to-peak) height difference of only ≈2 nm demonstrates that the film is smooth and uniformly covering the surface.

## SI 4 – C-axis out-of-plane texture of Sb and Sb₂Te₃ on amorphous substrates

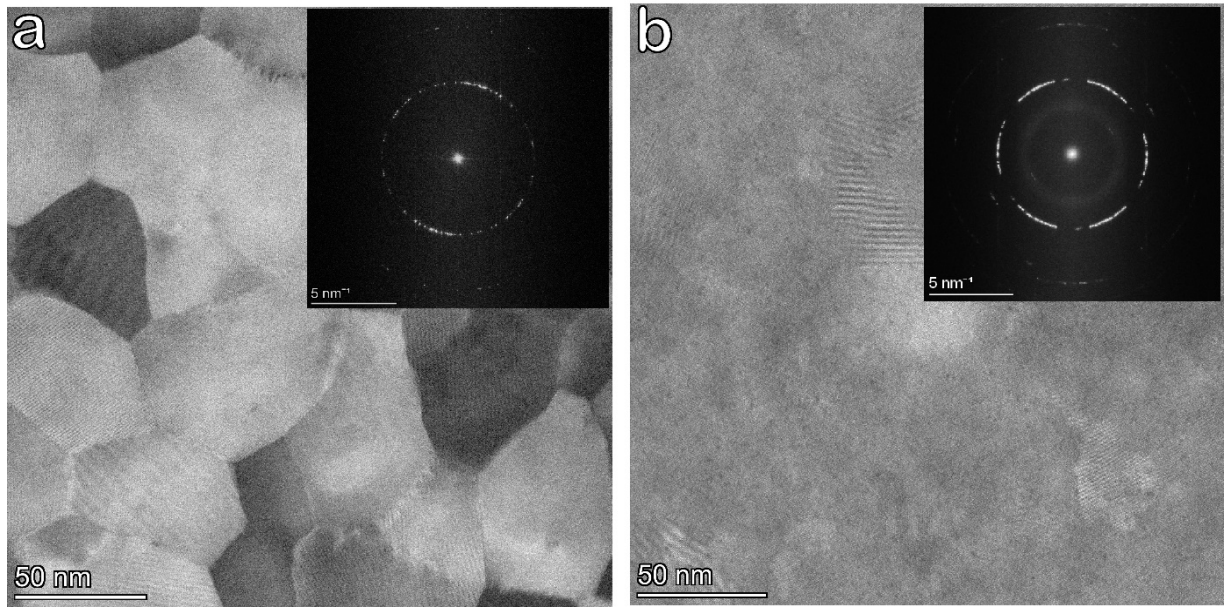

**Figure S5**. Plan-view (S)TEM images of crystal grains in Sb films with a thickness of ≈20 nm on (a) Sb$_2$Te$_3$ seed layer (heteroepitaxy) and (b) Sb seed layer (homoepitaxy) on a Si$_3$N$_4$ membrane. The insets show the FFT patterns. The rings in the FFT patterns indicate that the crystalline grains have a random in-plane orientation. On the other hand, the small numbers of rings indicate the grains are c-axis out-of-plane oriented, and some crystallographic orientations are not accessible from plan-view analysis.

Fig. S5a and S5b. show the TEM and STEM images of c-axis out-of-plane Sb films (≈20 nm thick) on Sb$_2$Te$_3$ and Sb seed layers, respectively. In both images, the insets show the respective FFT patterns. Fig. S5 captures several interesting results. The first one is that although we can see it visually from the plan-view images, the FFT patterns contain circular rings, indicators of randomly oriented grains. Of course, this is not surprising since we have shown proof that this is the case with RHEED globally and HAADF-STEM locally. One thing to notice in Fig. S5b is that the FFT patterns almost make a circular disk, unlike in Fig. 5 of the main text. The reason is that here, the field of view is relatively large compared to the field of view in Fig. 5. What the results indicate is that, although there are preferential in-plane orientations in smaller regions still containing multiple grains, in general, the grains are randomly oriented in large regions. Another point to note is that the FFT patterns only show a small number of rings. Since all grains are oriented with their c-axis out-of-plane, only the (000$l$) orientations are accessible from plan-view electron diffraction patterns. This is further evidence that there is a sharp c-axis out-of-plane texture.